\documentclass[aps,prA,twocolumn,groupedaddress]{revtex4}
\usepackage{graphicx}
\usepackage{setspace}
\usepackage{amsmath}
\usepackage{amssymb}
\usepackage{color}
\usepackage{array}
\usepackage{subfigure}
\usepackage{hyperref}
\usepackage{float}
\begin{document}
\title{Few-body ultracold reactions in a Bose-Fermi mixture}
\author{Chen Zhang, Javier von Stecher, and Chris H. Greene}
\affiliation{Department of Physics and JILA, University of Colorado, Boulder, Colorado 80309-0440, USA}
\date{\today}
\begin{abstract}
The spectrum of two bosons and two fermions in a trap is calculated using a correlated-Gaussian basis throughout the range of a broad Fano-Feshbach resonance. The calculations provide a few-body solution to the magneto-association of fermionic Feshbach molecules. This solution is used to study the time evolution of the system as the scattering length changes, mimicking experiments with Bose-Fermi mixtures near Fano-Feshbach resonances. The structure of avoided crossings in the few-body spectrum enables an interpretation of the dynamics of the system as a sequence of Landau-Zener transitions. The calculated molecule formation rate is compared with experimental observations.
\end{abstract}
\maketitle
\section{Introduction\label{Introduction}}
Cold atomic gases are of interest to physicists because the fundamental behavior of quantum matter can be investigated in such systems. After Bose-Einstein condensates \cite{Anderson14071995} and degenerate Fermi gases \cite{Regal2003} were created, Fano-Feshbach resonances were used to tune the interaction between atoms and to make molecules at ultracold temperatures. Since those early developments, these resonances have become essential for manipulating ultracold gases \cite{Bourdel2004,Jochim19122003}. Fano-Feshbach resonances in Bose-Fermi mixtures sparked research interest later \cite{Inouye2004,Goldwin2004}, because of the rich internal structure and the anisotropic interaction of heteronuclear Feshbach molecules. Loading fermionic heteronuclear molecules into optical lattices reduces their inelastic collisions and extends their lifetimes \cite{SOspelkaus2006,COspelkaus2006,Gunter2006,Best2009,Snoek2011,Olsen2009}. These experiments have motivated the development of theories to describe the dynamical processes of ultracold gases that occur during field ramps across a Fano-Feshbach resonance, which have been carried out from the perspectives of both many-body and few-body physics, for Bose-Einstein condensates and two-component Fermi gases \cite{Javier2007,Andreev2004}. However it has been challenging to describe the magneto-association of Feshbach molecules formed from a Bose-Fermi mixture.
\par
This article explores the spectrum and dynamics of three or four trapped atoms and shows how a few-body formulation can obtain accurate solutions of the system without making the standard approximations of many-body theory. This formulation provides an explicit representation of the avoided crossings responsible for interconversion between atomic states in a Bose-Fermi mixture and fermionic molecular states. Our results directly apply to optical lattice experiments with a few particles on one site, and could also be applied to larger Bose-Fermi mixtures through a frequency rescaling as in \cite{Borca2003,Javier2007}. The calculated dynamics provide a few-body perspective on the process of magneto-association of Feshbach molecules as varying degrees of adiabaticity and diabaticity.
\par
Specifically, to describe the behavior of an ultracold Bose-Fermi mixture from a few-body perspective in the vicinity of a Fano-Feshbach resonance, the spectrum is computed for two systems: one consisting of two bosons and one fermion (BBF) and the other consisting of two bosons and two fermions (BBFF). Although the one boson and two fermion system is also a subsystem of the four body system(BBFF), since it will not provide information of the trimer, we did not explore its property. In each system, interactions occur only through a short-range potential. To carry out this study for a concrete example, we take $^{87}\mbox{Rb}$ as the boson and $^{40}\mbox{K}$ as the fermion. The s-wave scattering length $\it{a}$ characterizing the short-range interaction between a boson and a fermion pair is tuned in the standard manner \cite{Greiner2003}. This model allows us to explore the magneto-association of fermionic molecules near a broad Fano-Feshbach resonance. Our solution of the few-body problem in this universal regime gives various properties accurately, including energy levels and the ramped time-dependent dynamics of the full quantum mechanical system. This type of analysis enables a detailed understanding of the global topology of the spectrum. The transition probabilities of this system into alternative possible final configurations are quantitatively predicted, for the case in which interactions change with time, as in field-ramp experiments \cite{Olsen2009}.
\par
\section{Method}
Eigenfunctions and eigenenergies for the few-body Hamiltonian [Eq.(\ref{Hamiltonian})] are determined using a variational calculation that utilizes a correlated-Gaussian basis \cite{Suzukibook}. This method has been previously used to treat the two-component fermion system \cite{Javier2007} and identical boson system \cite{Jose4boson2009}. After determination of the adiabatic spectrum, the time-dependent Schr\"odinger equation is solved numerically using the diabatic-by-sector method \cite{Javier2007, JavierThesis}. 
The few-body Hamiltonian adopted has only pairwise interactions between bosons and fermions in an isotropic harmonic trap:
\begin{equation}
 \mathcal{H}=\sum_{i\in B,F}\left(-\dfrac{\hbar^{2}}{2m_{i}}\nabla_{i}^{2}+\dfrac{1}{2}m_{i}\omega_{0}^{2}\textbf{r}_{i}^{2}\right)+
\sum_{i\in F}\sum_{j \in B}V(\textbf{r}_{ij}),
\label{Hamiltonian}
\end{equation}
where B denotes boson, F denotes fermion, and the interparticle potential is $V=V_{0}\exp(-\frac{r^{2}}{2d_{0}^{2}})$, i.e., an attractive Gaussian. The width $d_{0}$ of the Gaussian is fixed, and the depth $V_{0}$ is tuned to produce the desired two-body
 scattering length $\it{a}$. The boson-fermion reduced mass $\mu$, defined as $\mu=\frac{m_{B}m_{F}}{m_{B}+m_{F}}$, is half of the mass unit in our calculation. To obtain results independent of the details of the model potential, the range of the potential is limited to $d_{0} \ll a_{ho}$, where $a_{ho}=(\hbar/2 \mu \omega_{0})^{1/2}$ is
 the trap length based on this reduced mass. In the present calculation, $d_{0}=0.01 a_{ho}$ is used. The forms of the wave functions represented by this correlated Gaussian basis are listed in Table \ref{Gaussianbasis}.
\par
\begin{table*}
 \caption{Possible configurations and Gaussian basis functions. BBF denotes the trimer, BF denotes the dimer, and B (or F) denotes a single boson (fermion). $\textbf{R}_{cm}$ is the center-of-mass coordinate, 
$\psi_{0}$ is the ground state of the center-of-mass motion in a harmonic trap $\psi_{0}(\textbf{R}_{cm})=e^{-2R_{cm}^{2}/a_{ho}^{2}}$, 
and S denotes all the symmetrization or antisymmetrization appropriate for the identical bosonic or fermionic particles, $S=(1+\it{P}_{b_{i},b_{j}})(1-\it{P}_{f_{i},f_{j}})$, and P is the permutation operator.\label{Gaussianbasis}}
\begin{ruledtabular}
 \begin{tabular}{lll}
  System & configurations & basis function\\
\hline
BBF L = $0^{+}$ & BBF, BF+B, B+B+F & $\Psi_{d,u}(\textbf{r}_{1},\textbf{r}_{2},\textbf{r}_{3})=S\left\lbrace \psi_{0}(\textbf{R}_{cm})
e^{-\sum_{j>i}r_{ij}^{2}/2d_{ij}^{2}}\right\rbrace$ \\
BBFF L = $0^{+}$ & BBF+F, BF+B+F, B+B+F+F & $\Psi_{d,u}(\textbf{r}_{1},\textbf{r}_{2},\textbf{r}_{3},\textbf{r}_{4})=S\left\lbrace \psi_{0}(\textbf{R}_{cm})
e^{-\sum_{j>i}r_{ij}^{2}/2d_{ij}^{2}}\right\rbrace$ \\
BBFF L = $1^{-}$ & BBF+F, BF+BF, BF+B+F, B+B+F+F & $\Psi_{d,u}(\textbf{r}_{1},\textbf{r}_{2},\textbf{r}_{3},\textbf{r}_{4})=S\left\lbrace \psi_{0}(\textbf{R}_{cm})(\sum_{i}u_{i}z_{i})
e^{-\sum_{j>i}r_{ij}^{2}/2d_{ij}^{2}}\right\rbrace$ \\
 \end{tabular}
\end{ruledtabular}
\end{table*}
Each basis functions defined in Table \ref{Gaussianbasis} is characterized by a set of average interparticle distances $d_{ij}$ and by the orientations of Jacobi vectors $u_{i}$ (the $u_{i}$ are not needed for total angular momentum $L=0$ case). These entities are selected semi-randomly from several typical and intermediate configuration functions. Since both the long-range and short-range portion of the wave functions require an accurate description, the $d_{ij}$ range from a fraction of $d_{0}$ up to several $a_{ho}$. The typical size of the total basis set used in these calculations varies from 1000 to 5000. The advantage of the correlated-Gaussian basis set is that all the matrix elements can be evaluated analytically. One disadvantage of the correlated-Gaussian basis is that a large basis set may have a strong linear dependence, possibly resulting in numerical instability. Thus, we apply a method to suppress linear dependence during the generation and optimization process. The basis set is then fixed while $V_{0}$ is tuned to give different two-body scattering lengths. Matrix elements are calculated once and then used to obtain the spectrum throughout the entire range of the Fano-Feshbach resonance.
\par
The basis functions are tested for their convergence and accuracy in describing the system. The states in $\it{a} \sim \mbox{0}^{+}$ and $\it{a} \sim \mbox{0}^{-}$ limits are in agreement with the limiting-analytical behavior. The calculation elucidates the spectrum as a function of the scattering length and of the ramping dynamics. The inverse scattering length, which is the adiabatic parameter $\lambda$ of the ramping dynamics, is calculated in free space for the Gaussian potential by solving the equation of relative motion between two particles : $-\phi''(r)+(\frac{2m V_{0}}{\hbar^{2}}\exp(-\frac{r^{2}}{2d_{0}^{2}}))\phi(r)=0$. The zero-energy scattering solution generally has an asymptotic form equal to $\phi(r)\rightarrow C(r-a)$, where C is a constant and $\it{a}$ is the s-wave scattering length.
\par
With an optimized basis set, the spectra are calculated as a function of $1/\it{a}$, which presents a series of avoided crossings between the adiabatic levels near the point of trimer's formation and near unitarity($1/\it{a} \sim \mbox{0}$). Each avoided crossing can be qualitatively characterized by its width $\delta \lambda$, and by the range over which at least two adiabatic energy levels interact strongly, and grouped into two main categories: narrow crossings, whose $\delta \lambda \gtrsim 1/a_{ho}$, and wide crossings, whose $\delta \lambda \gg 1/a_{ho}$. Narrow crossings are approximated as exact, uncoupled crossings in the diabatization procedure, which gives a smooth, physical, partially diabatized spectrum as a function of $1/\it{a}$. Only the wide crossings are physically relevant for the range of ramping speeds of current interest, so the diabatization procedure leaves the wide crossings adiabatic.
\par
Thus the idea of diabatization is to compare the degree of similarity (quantitatively, the overlap) of the eigenfunctions at two nearby interaction strengths, by controlling the distance between the two points being compared, the best partially diabatized spectrum is selected. This selection is somewhat subjective, but after an appropriate diabatization, the structure of the avoided crossings permits a global qualitative view of the evolution of the system through different pathways from the weakly-interacting Bose-Fermi mixture at $\it{a}\rightarrow \mbox{0}^{-}$ to a strongly-interacting gas at $\it{a} \rightarrow \mbox{0}^{+}$. These pathways cross the universal region where $\it{a}$ goes from $-\infty$ to $\infty$, mimicking the experiments carried out in different laboratories \cite{Olsen2009,Shin2008}. Fig.\ref{Spectrum} presents the partially diabatic spectrum in the resonance region. 
\par
A more quantitative way to describe the avoided crossings is to evaluate the first derivative P-matrix coupling between the two energy eigenstates. The P-matrix is the nonadiabatic coupling between two adiabatic states, specifically: $P_{ij}=\left\langle \Psi_{i}| \frac{d\Psi_{j}}{d\lambda} \right\rangle$, where $\lambda$ is the adiabatic parameter. A narrow avoided crossing corresponds to a narrow and sharply peaked P-matrix centered at the crossing point, while a wide avoided crossings corresponds to a wide and smooth P-matrix. The diabatization procedure could also be carried out by calculating the P-matrix elements connecting each pair of adiabatic states and selecting those that are sufficiently wide according to some quantitative criterion to determine the adiabatic states. 
\par
After calculating the spectrum, the initial configuration is propagated using the time-dependent Hamiltonian. Starting from the ground state of the noninteracting limit ($\it{a} \rightarrow \mbox{0}^{-}$), the parameter $\lambda$ ($\lambda=\frac{1}{a_{sc}}$) is ramped through the resonance to the strongly interacting limit ($\it{a} \rightarrow \mbox{0}^{+}$) at different speeds $\nu=\frac{d \lambda}{d t}$. The Landau-Zener approximation is applied to interpret our results. The Landau-Zener approximation, predicts that the probability for a transition from the adiabatic $\Psi_{j}$ to $\Psi_{j}$ is $T_{ij}=\exp(-\chi_{ij}/\chi)$, where $\chi$ is the sweep rate of the adiabatic variable, and $\chi_{ij}$ is a parameter extracted from properties of the adiabatic eigenstates. 
\par
The nonadiabatic coupling (P-matrix) controls the probability of nonadiabatic transitions. Clark has shown that if a transitions has a form consistent with Landau-Zener approximation, then the P-matrix element for a transition from $\Psi_{i}$ to $\Psi_{j}$ has a Lorentzian form whose width(or height), along with the corresponding eigenenergies, characterize the Landau-Zener parameter $\chi_{ij}$ \cite{Clark1979}. Consequently, we evaluate all the potentially important $P_{ij}$ numerically and verify that the couplings between low-lying configurations generally have a smooth single-peak form that is approximately Lorentzian. However, the couplings between high-lying configurations do not have a perfect Lorentzian form, and can be multipeaked, as shown in Fig. \ref{PMatrix} and Table \ref{CompareP}. The most important $P_{ij}$ for this specific dynamical sweep corresponds to transitions among several configurations are listed in Table \ref{TimeEvolutionFunction}. The final probability distribution can be explained as a sequence of Landau-Zener transitions between these partially diabatic classes where the positions of the P-matrix elements' peaks determine a specific order in which the transitions occur. 
\par
Specifically for our system, the formation of the trimer happens at a negative scattering length, whereas the formation of the dimer(s) happens at unitarity. The Landau-Zener parameters obtained from the P-matrix analysis are listed in Table \ref{CompareP}. The Landau-Zener model shows qualitative agreement with the numerical results. Fig. \ref{Dynamics} presents the probability distribution of three or four particles system with total angular momentum 0 and 1 following a unidirectional ramp, as a function of the ramping speed $\frac{d\lambda}{dt}$. The numerical ramps are initiated at $\lambda_{i}\sim -8/a_{ho}$ and finalized at $\lambda_{f}\sim 8/a_{ho}$, where $d_{0} \ll a_{sc} \ll a_{ho}$ is satisfied, i.e. within the universal regime. As the speed is increased, the most probable final state changes from a bound trimer configuration to a bound dimer configuration and then to an atomic ground state, which is consistent with experimental observations.
\par
To connect our few body calculation to many body experiments, we construct a dimensionless ramping speed which has the density information of the system built in: $\chi=\frac{M}{\hbar \rho}|\frac{d\lambda}{dt}|$, in which $M$ is the total mass of the molecule, $\rho$ is the density of boson-fermion pair in the noninteracting system, and $\lambda=\frac{d(1/a_{sc})}{dt}$ is the ramping speed at which the inverse scattering length is ramped. Our calculation provides probability distribution into various final states across a Feshbach resonance as a function of the dimensionless ramping speed. At the same time, the dimensionless ramping speed could be extracted from experimental data at each value of $\frac{dB}{dt}$, our calculation predicts the molecule formation at that ramping speed. The above definition of dimensionless ramping speed is in a homogeneous system, however, we propose to use the peak density in an inhomogeneous system such as ultracold gases in harmonic trap. 
\par
In other words, the connection between the few body and many body system is realized by solving a few body problem in trap tight enough so that the peak density of the few body system is the same as the peak density of the many body system. The rescaled oscillator length depends on the peak density of the many body system $a_{ho}^{RS}=(\overline{\rho}/\rho_{exp})^{1/3}$, where $\overline{\rho}$ is the dimensionless peak density $\overline{\rho}=\rho*(a_{ho})^{3}$ from the few-body calculation, $\rho_{exp}$ is the peak density measured in experiments. For our calculation, the ratio of artificial trapping frequency to the experimental trapping frequency $\frac{\omega_{art}}{\omega_{exp}}$ is around 30. 
\par
In our calculation, the following units are used: $\hbar=1$, boson-fermion reduced mass $\mu=\frac{m_{B}m_{F}}{m_{B}+m_{f}}=0.5$, the mass ratio $\frac{m_{B}}{m_{F}}=\frac{87}{40}$, total mass of the molecule $M=m_{B}+m_{F}=2.31$, the trap frequency $\omega=1$, the oscillator length of the boson-fermion relative motion is $a_{ho}=(\frac{\hbar}{2\mu\omega})^{1/2}=1$, which is set to be the length unit, and the time unit $\frac{2\pi}{\omega}$. Density of molecules is defined as: $\int \rho d^{3}r=N$. For experimental system, N is around $10^{4}\sim10^{5}$ \cite{Olsen2009, OlsenThesis}, while for our calculation, N is 2. Dimensionless peak density $\rho=1.066/a_{ho}^{3}$.
\par
The artificially tight trap could be related to the number ratio of the few body system to the many body system. For experimental conditions, the temperature of the fermionic atoms is around $0.2 \sim 0.3T_{F}$, and the number of particles is around $10^4$ to $10^5$ \cite{Olsen2009, OlsenThesis}, so the density profile of the system could be described by the zero temperature Thomas-Fermi approximation \cite{SemiclassicalBook}. The peak density of a zero temperature noninteracting Fermi gas in an isotropic harmonic trap is proportional to $N^{1/2}$, when the particle number of N is much large than 1. So the relation between the experimental trapping frequency and the rescaled trapping frequency could be expressed in terms of the particle number ratio of few body calculation to many body experiments: the ratio $\frac{a_{ho}^{RS}}{a_{ho}^{exp}}\sim(\frac{2}{N})^{\frac{1}{6}}$. The temperature of bosons in the Bose-Fermi mixture is around $1.1\sim1.2 T_{c}$, thus the boson's peak density is tuned to match the peak density of the fermions in experiments. So the fermion peak density itself could give a good estimation of the maximum number of atom pairs near the center of the trap.
\par
\section{Results}
To relate our results to recent JILA experiments \cite{Olsen2009}, the Landau-Zener parameter for trimer-dimer-atom
 transitions $\delta=\chi_{mol}/\chi$ is cast in terms of experimentally accessible variables. Since the trimer and the dimer are not distinguished experimentally, we present one analysis for which the trimer formation is counted and another analysis where it is excluded from the count of molecules formed. 
\par
The adiabatic parameter in our calculation can be related to the ramping experiments, through an assumption that the dependence of the two-body scattering length $\it{a}(B)$ on the magnetic field is approximated in the usual manner as $a(B)=a_{bg}(1+\frac{w}{B-B_{0}})$, whereby $\frac{d (1/\it{a})}{dt}=(dB/dt)/(wa_{bg})$. Therefore, $\delta=\chi_{mol}(dB/dt)^{-1}\rho \hbar |w a_{bg}|/m$. The resonance is assumed to be sufficiently broad that, in the ramping range of magnetic field, $\omega \gg B-B_{0}$, whereby the ramping speed $\frac{d(1/a_{sc})}{dt}$ can be simplified to $\frac{1}{a_{bg}}\frac{1}{\omega}\frac{dB}{dt}$. This method allows us to interpret experimental data using $\beta(\frac{dB}{dt})$.
\par
The molecule fraction in the strongly interacting limit is the most relevant quantity to compare with experiments. For the present 4-body system, the molecular fraction is defined as the probability of ending up in the dimer-dimer configuration (this only occurs in the $L=1^{-}$ case) plus half of the probability of the BF+B+F configuration. If we count each trimer as a molecular bound state also, the trimer probability should be added.
\par
\subsection{Energy spectrum}
The energy spectrum is the first thing to calculate and it offers the launching point for further analysis. The systems selected for study are motivated by consideration of the relevant energy to the experimental temperature under the angular momentum and parity restriction. Since the temperature is in the ultracold regime, only low partial waves scattering can contribute. The fermionic nature of the dimer and trimer molecules formed implies that the ground state of the non-interacting system and the ground state of the strongly interacting system have different parities, so both parties are considered. 
\par
The BBF $L = 0^{+}$ system is a subsystem of the four body system (BBFF), and it is also computed because it provides the properties of the trimer. The BBFF $L = 0^{+}$ ground state is the ground state of the four-particle BBFF system in the $\it{a}\sim \mbox{0}^{+}$ limit. The BBFF $L = 1^{-}$ ground state is the ground state of the four-particle BBFF system in the $\it{a} \sim \mbox{0}^{-}$ limit. In this system a two-dimer configuration is allowed, thus it is possible to enhance the formation rate of the fermionic dimers compared to the BBFF $L=0^{+}$ state. Because the dimers formed are identical fermions, the relative angular momentum between two $s$-wave dimers must be odd, which makes states with nonzero total angular momentum state possibly important. This is different from the bosonic dimer case \cite{Papp2006, Inouye2004} 
\par
\begin{figure*}
\begin{tabular}{l}
\includegraphics[width=18cm]{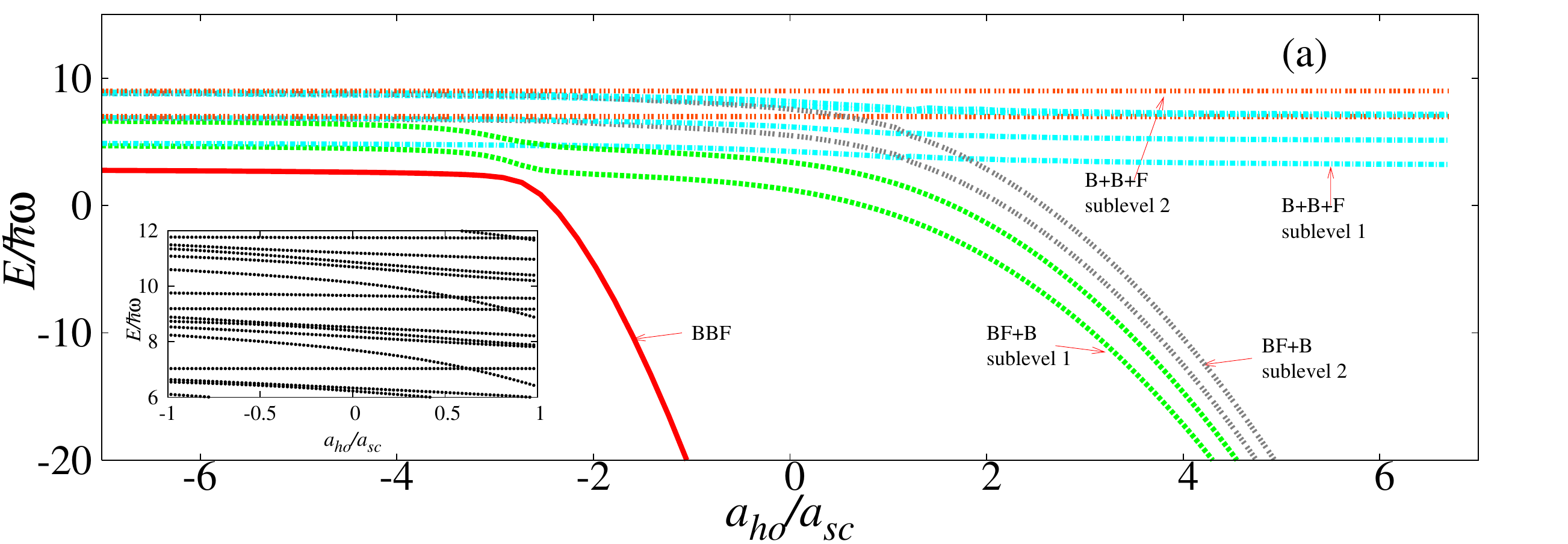} \\
\includegraphics[width=18cm]{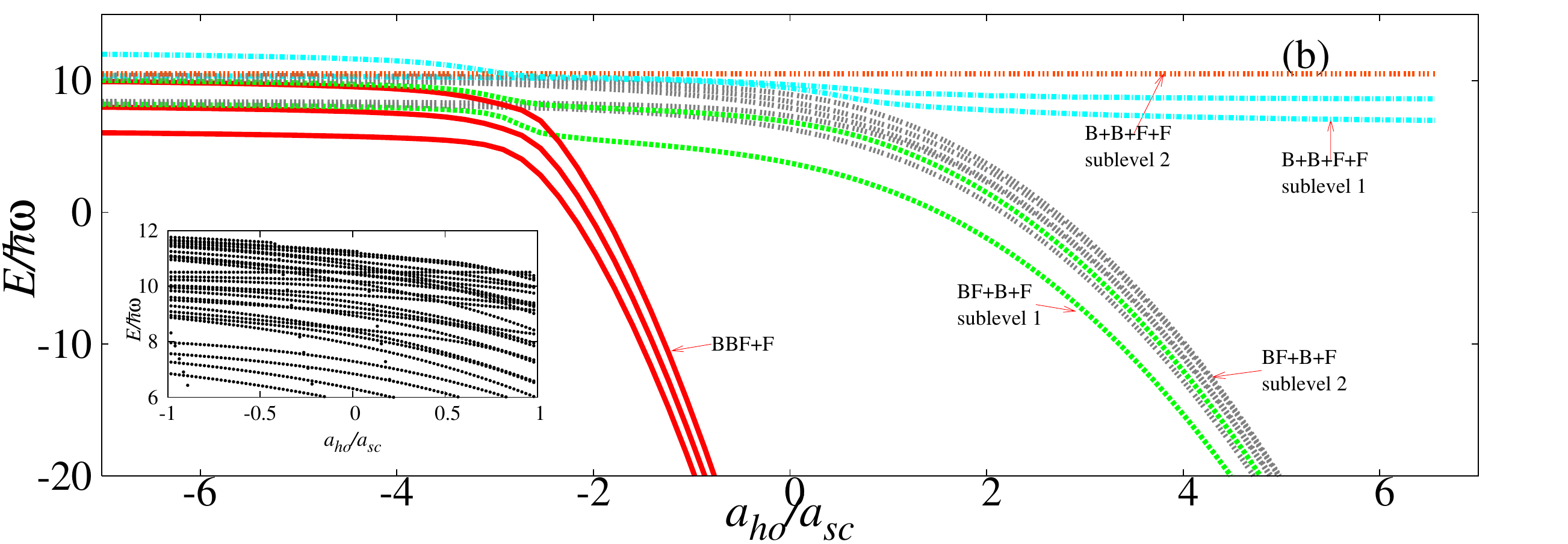} \\
\includegraphics[width=18cm]{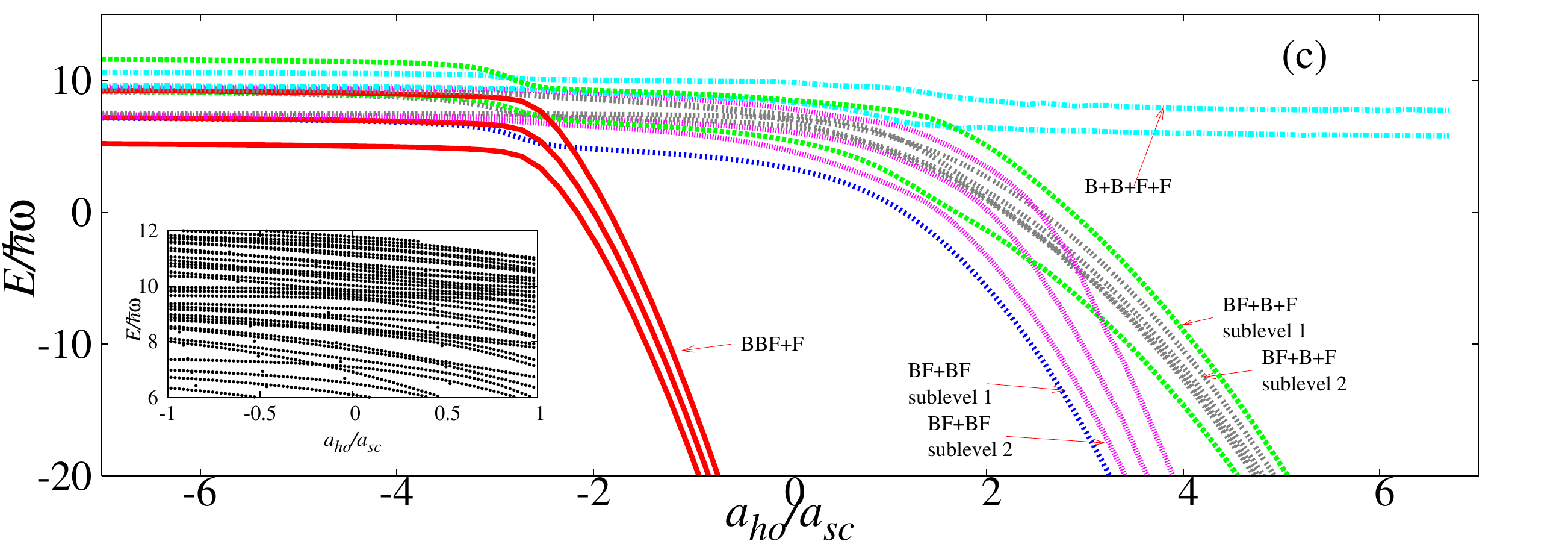}
\end{tabular}
\linespread{1} 
\caption{The energy spectra are plotted as functions of the dimensionless inverse scattering length in trap oscillator units $a_{ho}/a_{sc}$ for the three few-body systems considered. Classes and typical sublevels in each class are pointed by arrows on the figure. (a) BBF $L=0^{+}$. (b) BBFF $L=0^{+}$. (c) BBFF $L=1^{-}$. The inset shows avoided crossings and a pictorial demonstration of our diabatization near unitarity. \label{Spectrum}}
\end{figure*}

In each calculation, the various classes of final states are distinguished by their configuration in the $\it{a}\sim \mbox{0}^{+}$ limit. Both the energy and the pair-correlation function are examined to determine the state configuration. The sublevels within each class have different curvatures near avoided crossings. The ``sublevels'' that are close in energy just before reaching the crossing region are often degenerate states in the noninteracting limit. However the boson-fermion pair in these states have differences in their relative motion, e.g., different angular momentum or a different number of radial nodes; consequently the ramp affects these states differently, especially near the avoided crossings. In the $\it{a} \sim \mbox{0}^{+}$ regime, levels that may not be energetically close in the noninteracting limit, can sometimes end up in the same configuration, analogous to the promotion of diatomic electronic orbitals in the united atom versus the separated atom limits. All of the present calculations are performed at $\omega=1$, $a_{ho}=1$, $m_{B}=1.5815$, $m_{F}=0.72988$ ($^{40} \mbox{K}$ and $^{87} \mbox{Rb}$). Figure \ref{Spectrum} presents the partially-diabatized energy spectrum for the few-body systems considered here.
\par
\subsection{Time evolution}
The time-evolution of the initial state correlates with the adiabatic energy spectrum, e.g., when the relative probabilities change rapidly at some time $t_{0}$, there is usually a corresponding avoided crossing in the energy spectrum at $t_{0}$. But in the weak coupling regime, the probability in each class remains steady and the phase of the probability amplitude is described by the WKB-type evolution. 
\par
\begin{figure*}
 \centering
\begin{tabular}{lll}
\includegraphics[width=6cm,height=4cm]{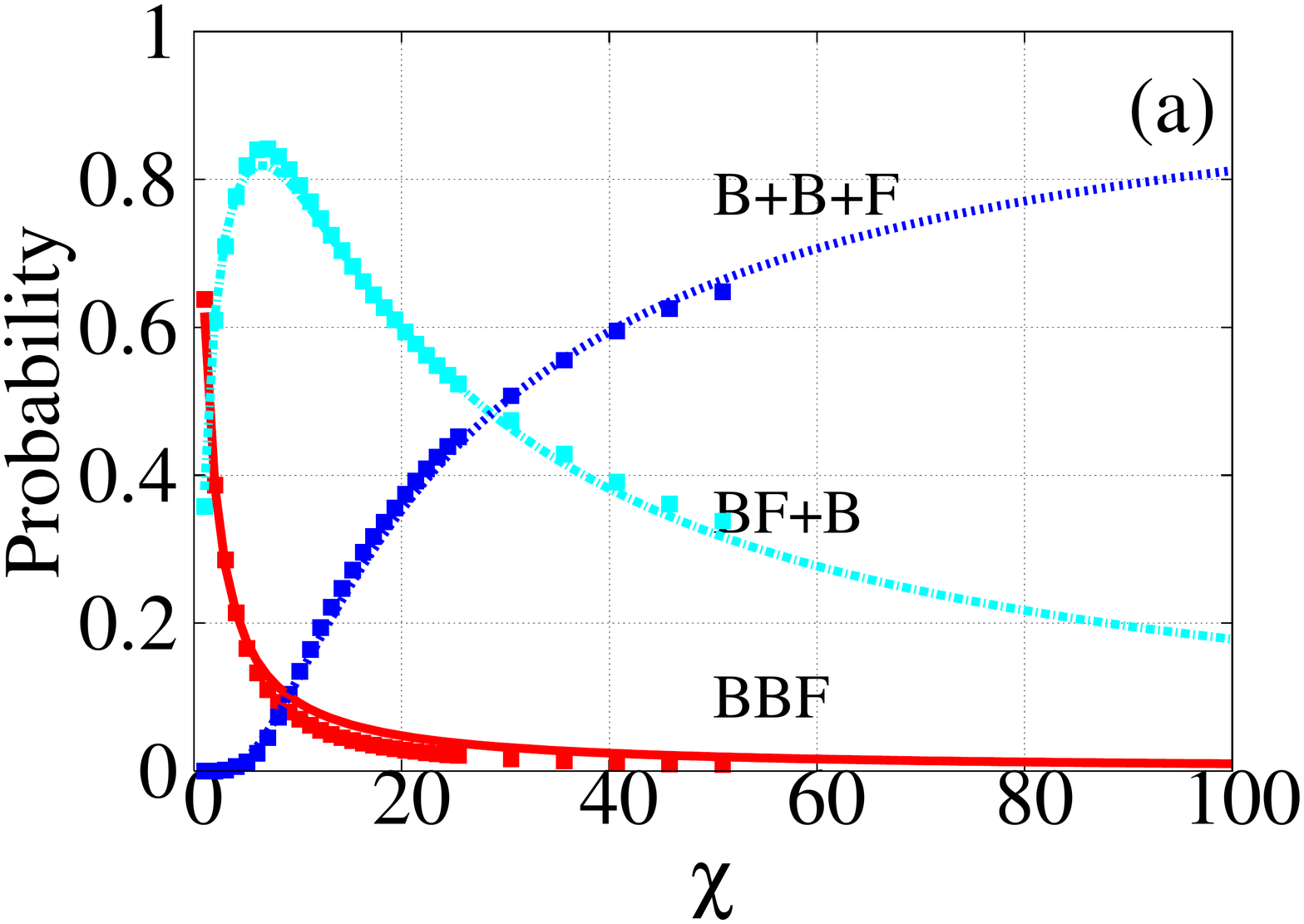} &
\includegraphics[width=6cm,height=4cm]{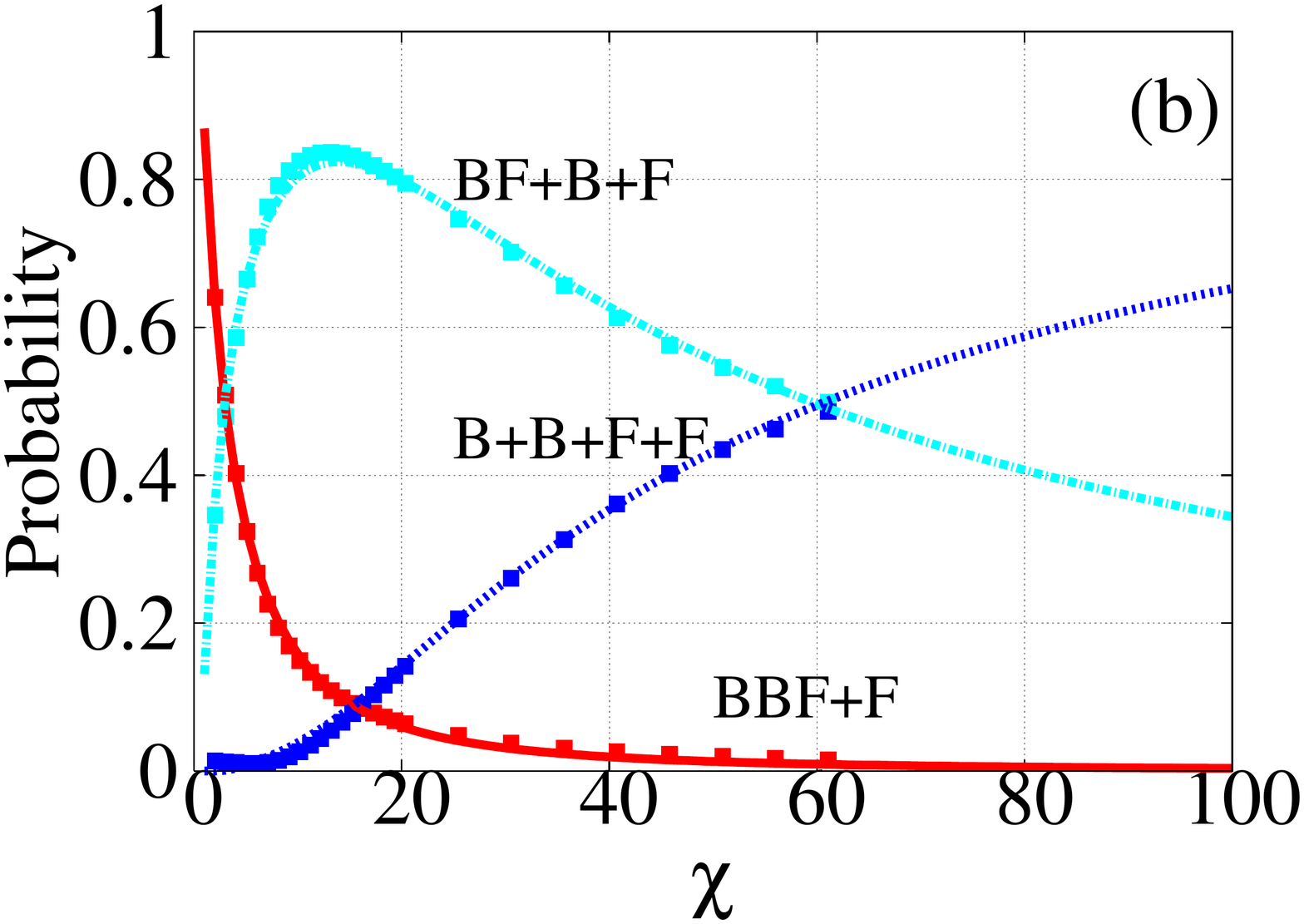} & 
\includegraphics[width=6cm,height=4cm]{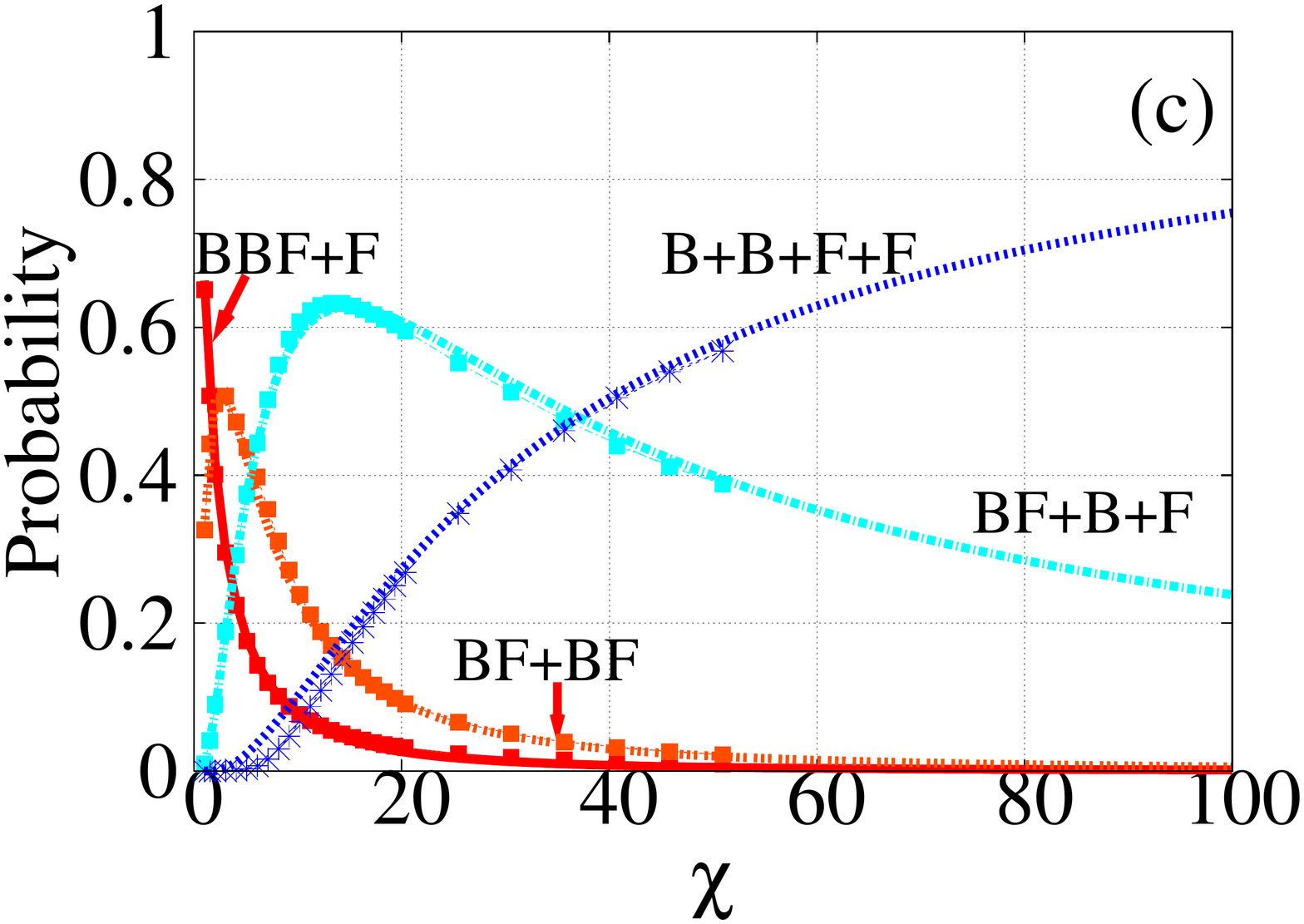}
\end{tabular}
\linespread{1} 
\caption{The final probability distribution are plotted versus the ramping speed. Starting from the ground state of the nearly noninteracting fewbody system, as the dimensionless ramping speed $\chi=\frac{M}{\hbar\rho}|\frac{d\lambda}{dt}|$ ($\lambda=|\frac{d(1/a_{sc})}{dt}|$) is varied, one can access various strongly interacting final states. On these figures, crosses $+$ give the result of full numerical calculations, while the curves are our best fits using a Landau-Zener model. The functional forms for each fit are given in Table \ref{TimeEvolutionFunction}. (a) BBF $L=0^{+}$. (b) BBFF $L = 0^{+}$. (c) BBFF $L = 1^{-}$. Curves are fit from numerical result and ``+'' is numerical calculation. \label{Dynamics}}
\end{figure*}

The final probability distribution predicted by the numerical time evolution is fitted to the Landau-Zener frame work by including the minimum number of necessary transitions. The two main criteria for ranking the importance of any given transition are (1) the initial population of the relevant states, indicating the population that can be transferred, (i.e., if the initial population of two states are both small, even if the coupling between these two levels are strong, this transition's contribution to the final probability distribution is still minor) and (2) the Landau-Zener parameter given by our computed P-matrix analysis, characterizing the strength of the transition and its corresponding characteristic ramp rate. The optimal fitting result is shown in Fig. \ref{Dynamics} and Table \ref{TimeEvolutionFunction}. In general, as the energy spectrum gets more complicated, the minimum number of transitions needed to get a qualitatively good fit to the numerical time-evolution increases.
\par
The fit shows that our numerical results appear to be well described by an incoherent Landau-Zener model. However, in principle, to analyze the interference of transition amplitude pathways leading to indistinguishable final states, the phase information must be included, as in the Stueckelberg case \cite{Larson2000}, where inclusion of the phase information was essential to determine the oscillation of probabilities as functions of energy. In our calculation, the incoherent Landau-Zener model fits adequately, as was the case for the two-component fermion system \cite{Javier2007}. One possible reason is that the initial population for each important transition has always been chosen to reside in a single state, which results in little if any interference. Some evidence exists to support this suspicion; specifically, when the initial state chosen for the time evolution is picked to be the ground state, a probability oscillation is observed in some cases as a function of the ramping speed. Moreover, when our few-body calculation are applied to the many-body gas through a frequency rescaling, one might expect any any phase coherence at the few-body level could be washed out for the many body system, since there are multiple sources of decoherence in a many-body gas.
\par
\begin{table}
\caption{The final probability distribution of time evolution versus ramping speed is represented as a sequence of Landau-Zener transitions. $T_{ij}$ is defined as $\exp(-\chi_{ij}/\chi)$, where $\chi_{ij}$ is the fitted Landau-Zener parameter, and $\chi$ is the ramping speed $\frac{d\lambda}{dt}$, $\lambda=\frac{a_{ho}}{a_{sc}}$. The fitted $\chi_{ij}$s are listed in Table\ref{CompareP}. \label{TimeEvolutionFunction}}
\begin{ruledtabular}
\begin{tabular}{ll}
\multicolumn{2}{l}{\tiny{BBF L = 0}}\\
\hline
\tiny{$p_{1}$ (trimer)}  & \tiny{$1-T_{12}$}\\
\tiny{$p_{2}$ (dimer + 1 atom)} & \tiny{$T_{12}(1-T_{23})$}\\
\tiny{$p_{3}$ (3 atoms)}  & \tiny{$T_{12}T_{23}$}\\
\hline
\multicolumn{2}{l}{\tiny{BBFF L = 0}}\\
\hline
\tiny{$p_{1}$(trimer)} &  \tiny{$(1-T_{12})(1-T_{13})$} \\
\tiny{$p_{2}$ (dimer + 2 atoms)} & \tiny{$T_{12}(1-T_{23})$} \\
\tiny{$p_{3}$ (4 atoms )} & \tiny{$T_{12}T_{23}+(1-T_{12})T_{13}$} \\
\hline
\multicolumn{2}{l}{\tiny{BBFF L = 1}}\\
\hline
\tiny{$p_{1}$ (trimer + 1 atom)} & \tiny{$(1-T_{12})(1-T_{13})$} \\
\tiny{$p_{2}$ (2 dimers)}  & \tiny{$T_{12}(1-T_{23})(1-T_{24})$} \\
\tiny{$p_{3}$ (1 dimer + 2 atoms)}  & \tiny{$(T_{12}T_{23}+(1-T_{12})T_{13})(1-T_{34})$} \\
\tiny{$p_{4}$ (4 atoms)} & \tiny{$T_{12}T_{23}T_{34}+(1-T_{12})T_{13}T_{34}+T_{12}(1-T_{23})T_{24}$} \\
\end{tabular}
\end{ruledtabular}
\end{table}
\subsection{P-matrix analysis}
The Landau-Zener parameters can alternatively be extracted directly from the numerically calculated nonadiabatic coupling P-matrix, as an alternative to the fitting procedure discussed above. Each P-matrix element connecting two low-lying states corresponds to a single important transition between state $\Psi_{i}$ and $\Psi_{j}$, where the Landau-Zener parameter $\chi_{ij}$ can be extracted without ambiguity. The reasonable agreement with the fitted Landau-Zener model parameters suggest that those transitions are adequately described by a two-level model, and that contributions from other levels' nonadiabatic coupling to other levels in the transition are negligible. The dominant pathways are apparently isolated from others sufficiently far to justify this approximation. However, some P-matrix elements involving high-lying states have multiple-peaks, non-Lorentzian shapes for each peak, and apparent discontinuities in the explored region of $a_{ho}/a_{sc}$. This phenomena could have several origins, the basic reason for its violation of Landau-Zener model validity is the non-isolated multi-channel nature of the transition, i.e. near the transition point, the nonadiabatic coupling between more than 2 levels at a time could be non-negligible. The following methods are used to address these issues.
\par
\begin{figure}
\centering
\begin{tabular}{l}
\includegraphics[width=8.5cm]{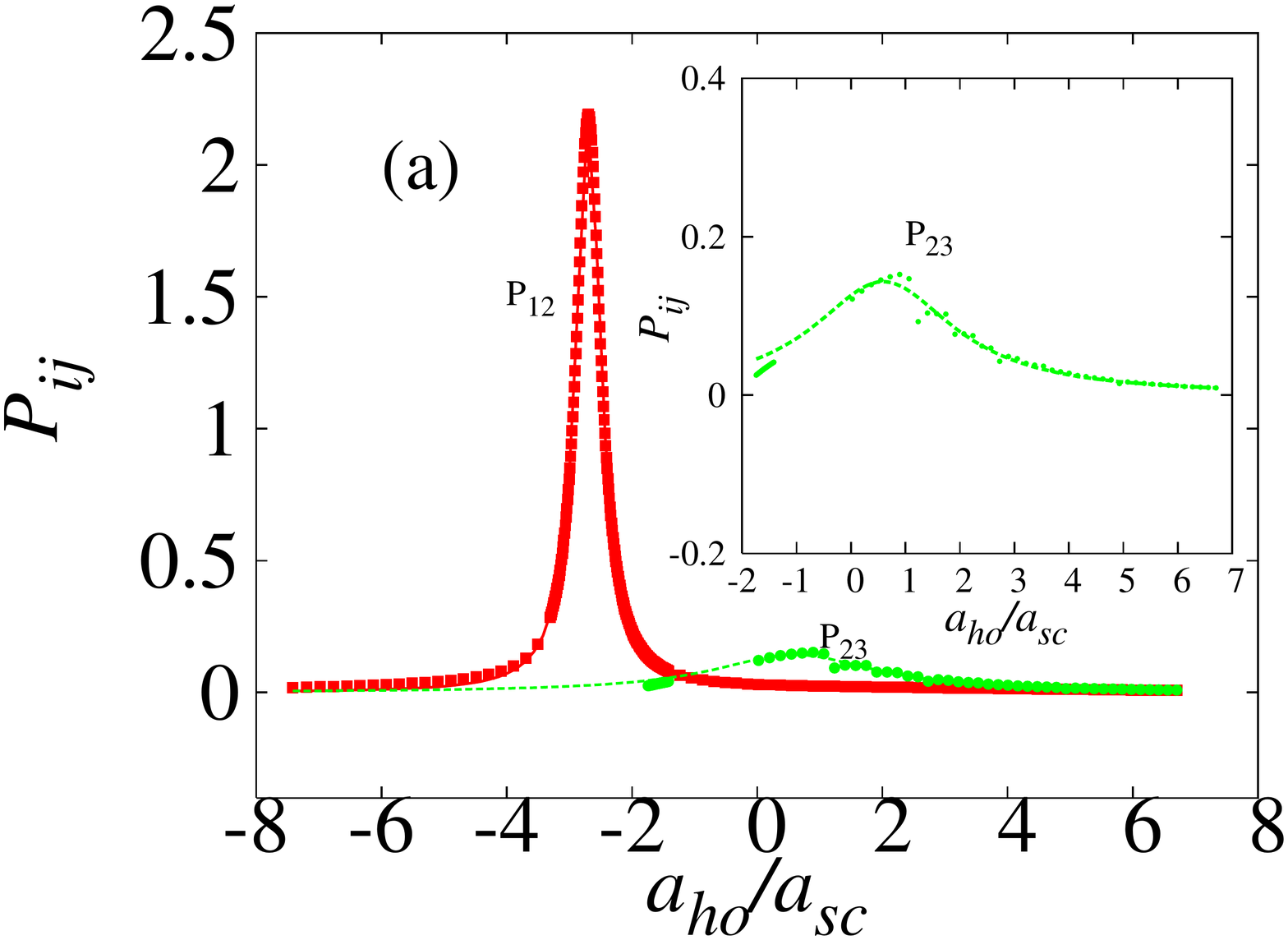}\\
\includegraphics[width=8.5cm]{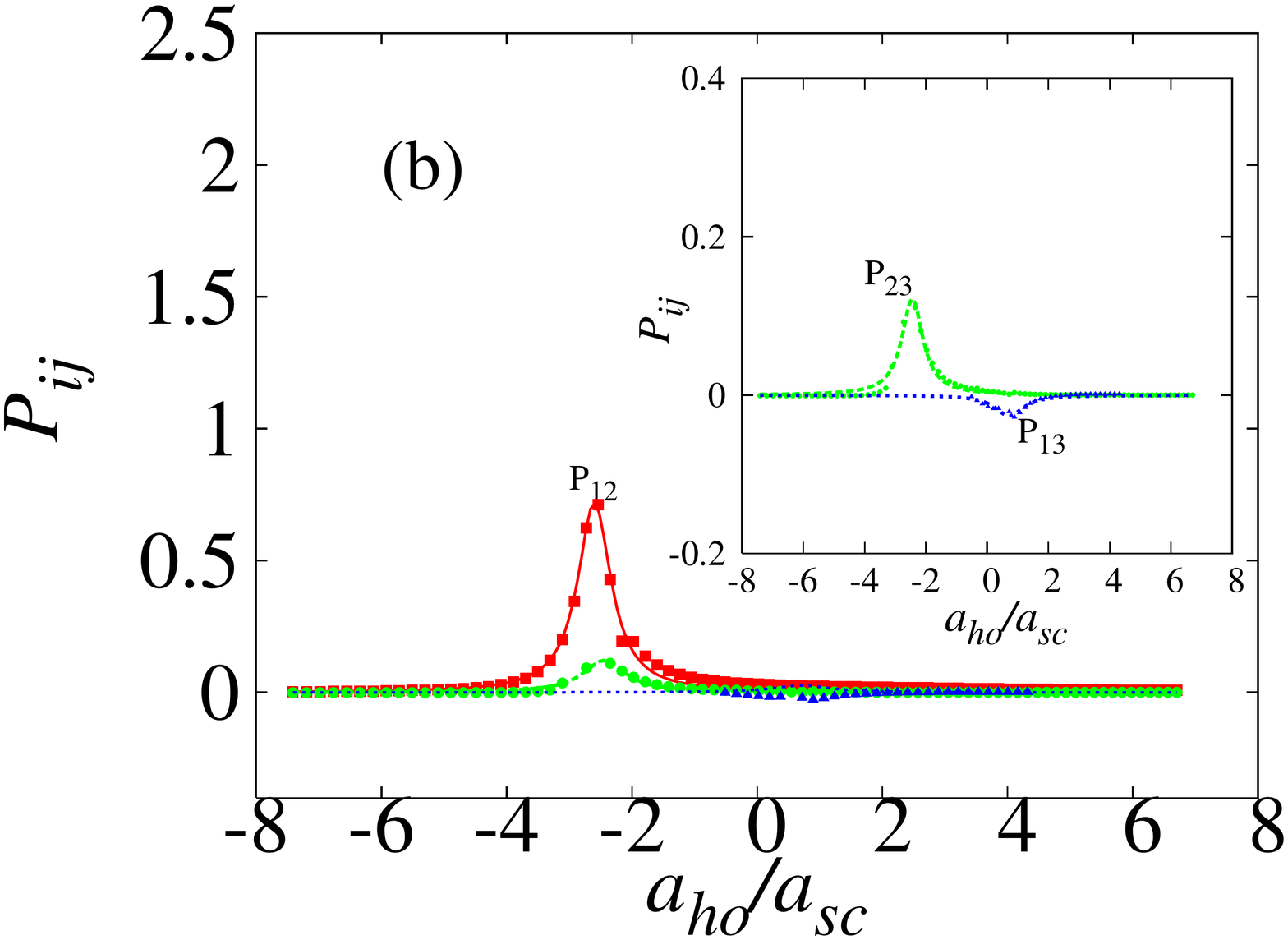}\\
\includegraphics[width=8.5cm]{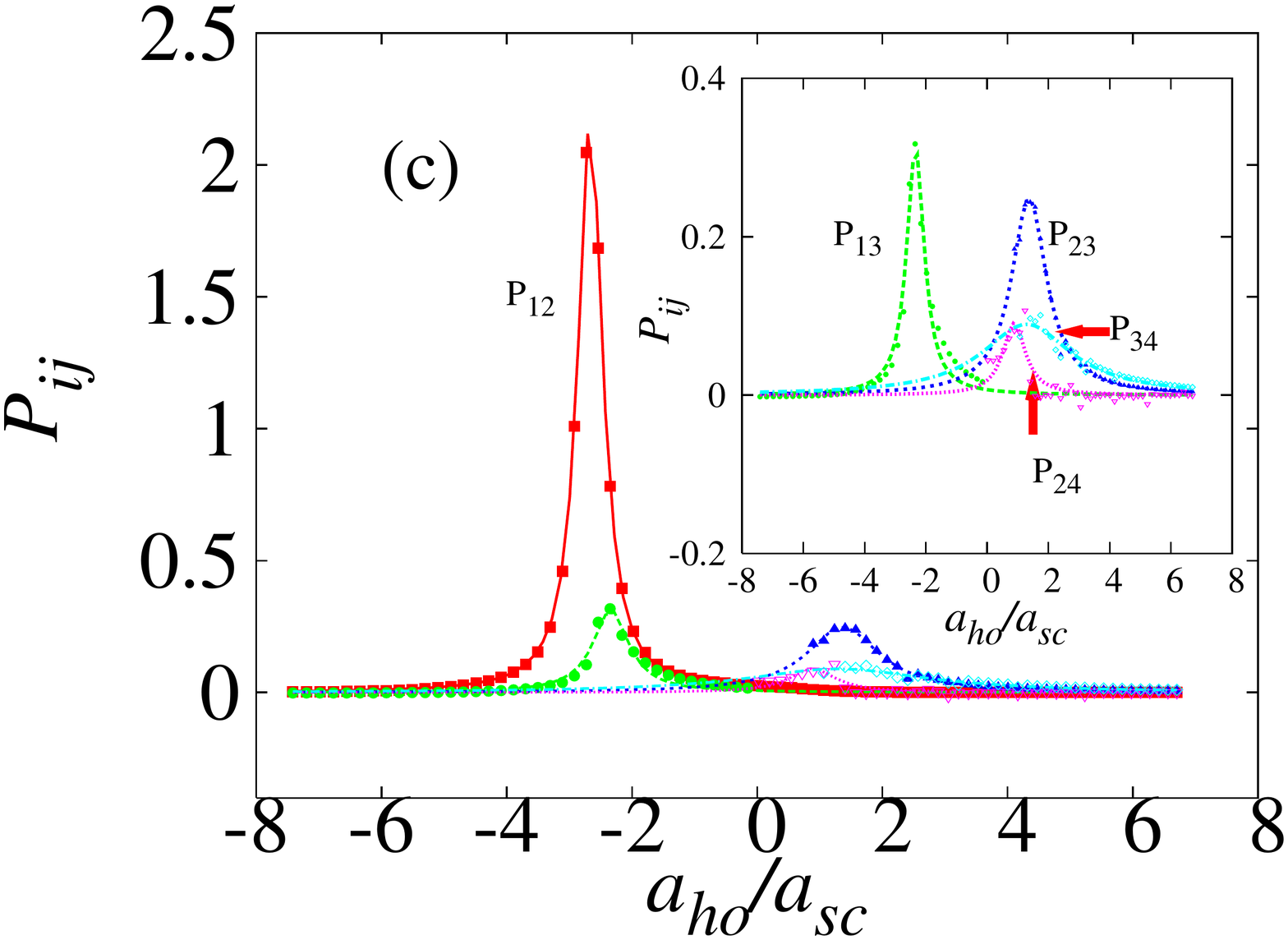}
\end{tabular}
\caption{Some important P-Matrix elements calculated as $P_{ij}=\left<\Psi_{i}|\frac{d}{d\lambda}\Psi_{j}\right>$, in which $\lambda=\frac{a_{ho}}{a_{sc}}$. (a) BBF $L = 0^{+}$. (b) BBFF $L = 0^{+}$.(c) BBFF $L = 1^{+}$. The inset of each figure contains a zoom in of the P-matrix elements except for $P_{12}$. Both numerical calculated $P$-matrix elements and their fitted curve into Lorentzian form are shown. The number labeling of $P$-matrix ($P_{ij}$) between different classes are: (a) 1:BBF, 2: BF+B, 3: B+B+F, (b) 1: BBF+F, 2: BF+B+F, 3: B+B+F, (c) 1: BBF+F, 2: BF+BF, 3: BF+B+F, 4: B+B+F+F.  \label{PMatrix} }
\end{figure}

For the case of multiple-peaks in a give P-matrix element, the most important peak can be selected by considering both the order and strength of the transition. For a non-Lorentzian shape, the most important peak is fitted as a Lorentzian form with two fitting parameters instead of using just one parameter as in the standard Landau-Zener model as reformulated by Clark \cite{Clark1979}. The P-matrix elements for the most important transitions are shown in Fig. \ref{PMatrix}, along with the best fits and the full numerical calculation. In a standard Landau-Zener model, the predictions extracted from the height and width of the P-matrix should coincide, but in our calculation not all the P-matrix elements have this feature because some exhibit non-Lorentzian shapes. For instance, consider $P_{13}$ in the BBF system, for which the fitted Landau-Zener parameters extracted from our numerical time-evolution are seen to have values close to at least one of the two fitted P-matrix parameters or, they fall between the two alternative values fitted to either the height or the width of the assumed Lorentzian form of the $P_{ij}$.
\par 
The trimer-dimer transition, corresponding to $P_{12}$, is sharper than the atom-dimer transition, which suggests that adiabatic formation of trimer states requires a slower ramping speed than the formation of the dimer states. This fact is consistent with the result of our numerical time propagation. The Landau-Zener parameter characterizes quantitatively the typical ramping speed needed in order to stay in a trimer state or to ramp into either dimer states or atomic states.
\par 
\begin{table}
 \caption{Comparison of Landau-Zener parameters extracted directly from the nonadiabatic coupling P-Matrix calculation and from fitting the Landau-Zener model directly to our numerical time evolution. Here the terms height and width mean that the Landau-Zener parameters $\chi_{ij}$ below have been fitted to the height and width of the corresponding $P$-matrix elements. The incoherent transition functional form is listed in Table \ref{TimeEvolutionFunction}. The number labeling of $\chi_{ij}$ is consistant with the labeling rule in Fig. \ref{PMatrix}. \label{CompareP}}
\begin{ruledtabular}
\begin{tabular}{llll}
 $\chi$  & P-matrix & P-matrix  & time-evolution\\
 & (width) & (height)  & \\
\hline
\multicolumn{3}{l}{BBF $L = 0^{+}$}\\
\hline
$\chi_{12}$ & 1.18 & 1.12 &  0.98\\
$\chi_{23}$ & 22.4 & 47.8 &  19.9\\
\hline
\multicolumn{3}{l}{BBFF $L = 0^{+}$}\\
\hline
$\chi_{12}$ & 1.57 & 1.89  & 2.06\\
$\chi_{13}$ & 11.7 & 110  & 19.1\\
$\chi_{23}$ & 11.8 & 371  & 43.3\\
\hline
\multicolumn{3}{l}{BBFF $L = 1^{-}$}\\
\hline
$\chi_{12}$ & 1.23 & 1.21 & 1.09\\
$\chi_{23}$ & 6.15 & 76.5  & 4.07 \\
$\chi_{13}$ & 6.25 & 28.8  & 13.3\\
$\chi_{24}$ & 3.94 & 48.4  & 15.3\\
$\chi_{34}$ & 29.7 & 92.5  & 28.5\\
\end{tabular}
\end{ruledtabular}
\end{table}
\subsection{Molecule formation percentage}
\begin{figure}
\centering
\begin{tabular}{ll}
\includegraphics[width=4cm]{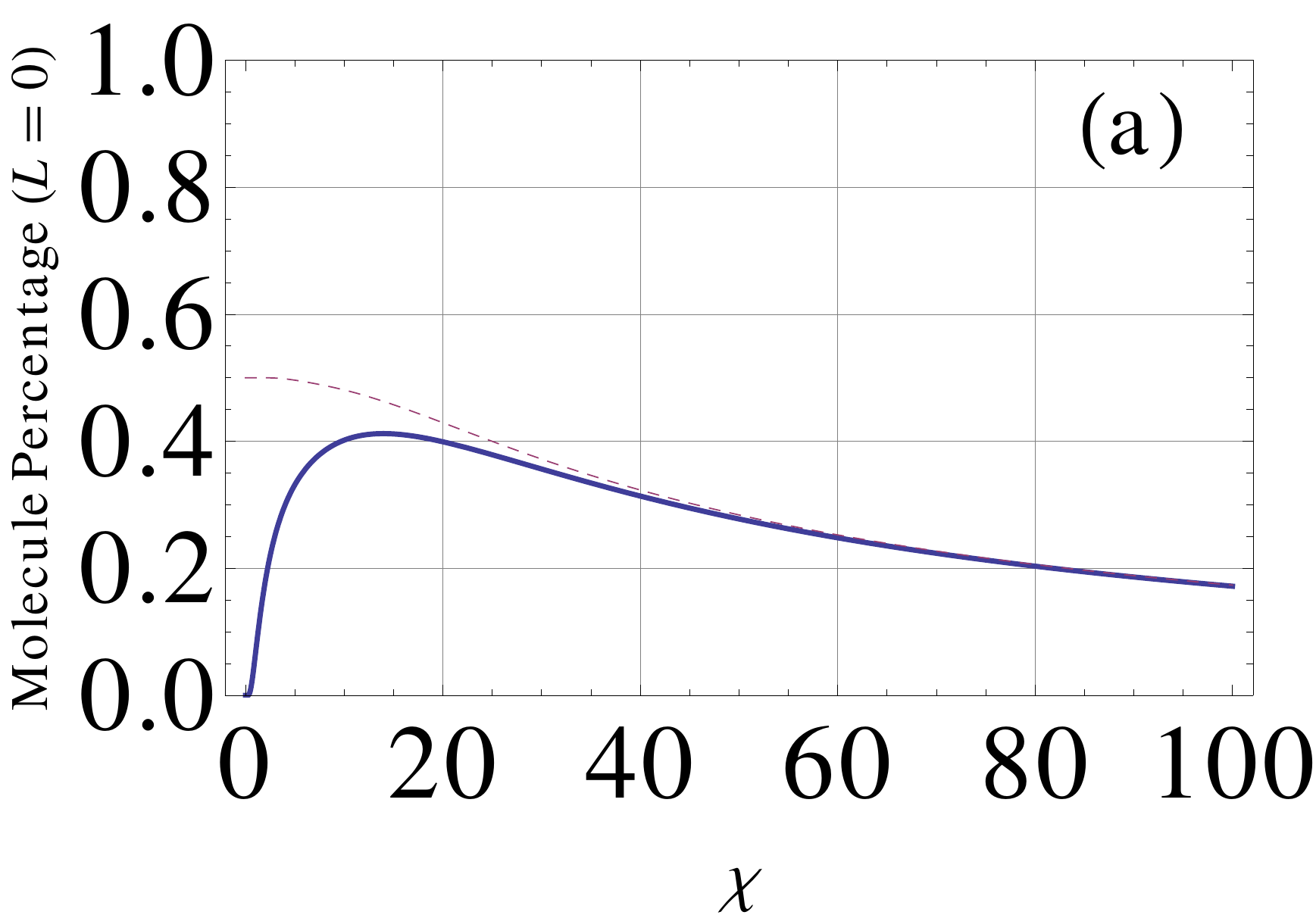}&
\includegraphics[width=4cm]{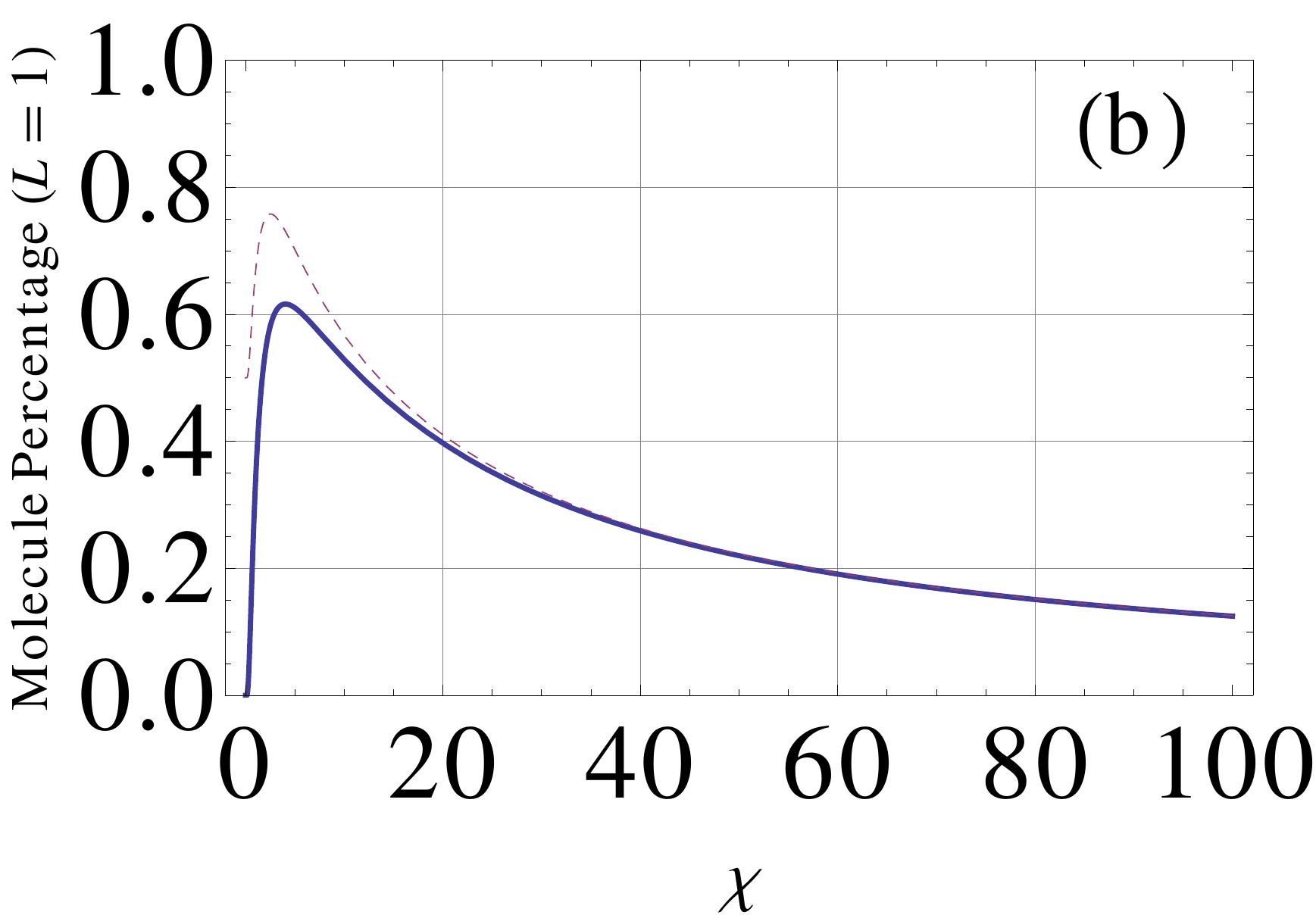}
\end{tabular}
\linespread{1} 
\caption{The molecule formation percentage is depicted as a function of the dimensionless ramp speed for (a) $L=0^{+}$ and (b) $L=1^{-}$ BBFF systems. Since the trimer and dimer were not distinguished experimentally, two separate analysis are presented here. In one analysis, a trimer is counted as a molecule (dashed) and in the other analysis the trimer formed is excluded from the molecules counted (solid). \label{MoleRatio}}
\end{figure}

Prediction of the molecule formation ratio as a function of the inverse scattering length ramping speed is the strongest connection between our few-body calculation and the experimental many-body observations. Fig. \ref{MoleRatio} shows the molecule formation ratio in BBFF $L=0^{+}$ and $L=1^{-}$ systems. Although the $L=1^{-}$ system has a higher upper limit for its molecule formation ratio, the formation rates can hardly be distinguished at fast ramping speed. 
\par 
Various experiments have been carried out that explored the maximum formation rate of heteronuclear Feshbach molecules in recent years, including $^{6}\mbox{Li} ^{7}\mbox{Li}$ \cite{Truscott30032001,Schreck2001}, $^{6}\mbox{Li}^{23}\mbox{Na}$ \cite{Hadzibabic2002}, $^{40}\mbox{K} ^{87}\mbox{Rb}$ \cite{Roati2002,Modugno27092002,Olsen2009} and $^{6}\mbox{Li} ^{133}\mbox{Cs}$ \cite{Mudrich2002}. The highest conversion through adiabatic magnetic field ramping across Fano-Feshbach resonance in $^{40}\mbox{K} ^{87}\mbox{Rb}$ given by JILA \cite{Olsen2009,Goldwin2004} is around 36 percent of the minority ($^{87}\mbox{Rb}$). Our calculation gives the conversion rate as function of ramping speed in Fig.\ref{MoleRatio}. By comparing the Bose-Fermi mixture experiment to former experiments, we notice that in Bose-Fermi mixtures the conversion rate never approach anywhere close to unity. The reason for this low conversion efficiency compared to the two-component fermion system and ultracold boson system \cite{Hodby2005,Papp2006,Goldwin2004}, which are described well by a stochastic phase-space paring model \cite{Hodby2005}, however this model is under debate. The predicted molecule formation rate in the stochastic phase-space pairing model is in general higher than the observed rate in magneto-association experiments \cite{Olsen2009}. In our few-body calculation, the possibility of trimer formation provides a way to potentially understand why the observed molecule formation rate might be reduced: the trimer is more deeply bound than the Feshbach molecules, and might decay rapidly into one deeply bound dimer if struck by another atom, or if it predissociates, in which case the ``spin flip'' method for detecting the dimer \cite{Olsen2009} would not be expected to detect a trimer bound state.  
\par
Our few-body calculation could be directly applied to understand optical lattice experiments when the tunneling between sites is small. Ideally, one boson and one fermion in one site is expected, and the presence of additional bosons is likely to introduce three-body loss. However, our calculation suggests that within an appropriate sweeping speed range, more than 2 particles on a single site could be directly converted into a molecular bound state. 
\section{Discussion}
The Landau-Zener model yields a qualitative picture of the dynamical features of the trapped few-body system. It also offers some quantitative predictions for the transition probabilities and the characteristic range of ramp speeds that cause a change-over from a high rate of molecule formation to a low rate. However, the question of whether a Landau-Zener function is the appropriate functional form to describe the molecule formation fraction in a large system is still under debate \cite{Pazy2006,Chen2011}. The Landau-Zener model (LZ) for three or four particles does not predict a single LZ function but rather a combination of different LZ terms that incoherently add up. Owing to its formulation in terms of the adiabatic eigenfunctions, the Landau-Zener model could identify each important transition into various possible final configurations reasonably well. Thus the ``sequence of transitions'' functional form appears to be consistent with our numerical simulations. But a two-level Landau-Zener model could not readily incorporate the effect of other nearby levels that may couple to one of the two levels, while the final probability is regarded as a sum of the probabilities for transitions into all levels in each class, which is a limitation of the Landau-Zener model for analyzing our numerical results.
\par
\begin{figure}
 \centering
\begin{tabular}{ll}
\includegraphics[width=4cm]{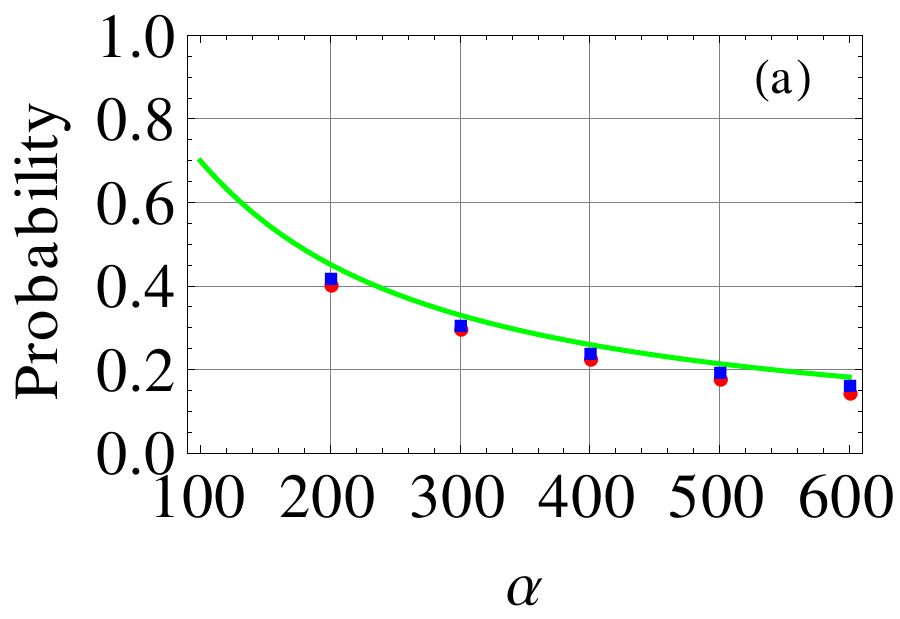}&
\includegraphics[width=4cm]{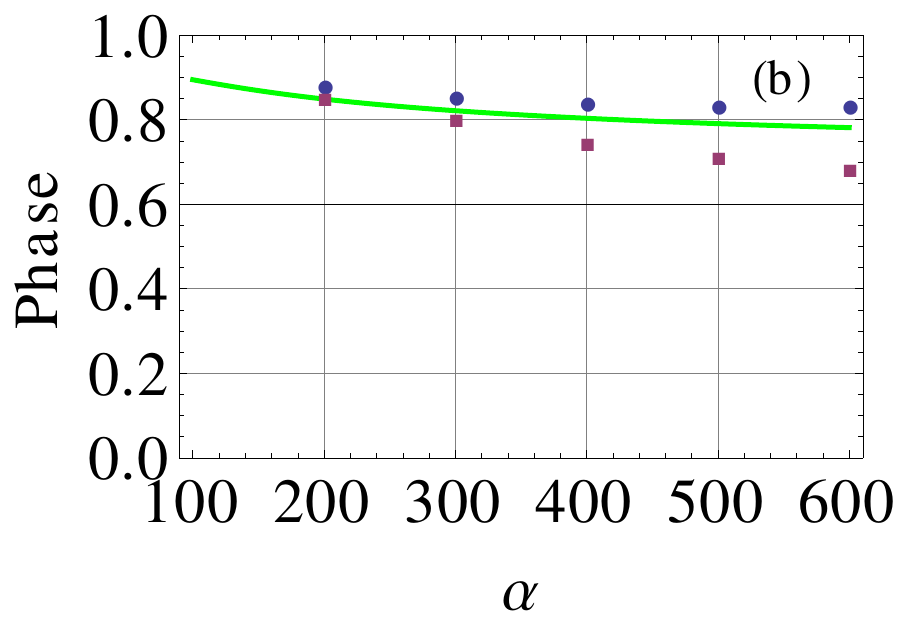}
\end{tabular}
\linespread{1} 
\caption{Comparison of our numerical calculation carried out in the adiabatic representation of a two-level system (squares) with the numerical calculation in the diabatic representation for a multilevel system (diamonds), and the analytical Landau-Zener model's prediction for the asymptotic behavior of the transition probability (solid line). (a) Probability to form a trimer state as a function of ramping speed $\alpha=\frac{dV_{0}}{dt}$ in the two-level adiabatic calculation. The levels in included are the lowest BBF+F channel and lowest BF+B+F channel. (b) Cosine function of the phase accumulated during the transition, where the phase accumulation is defined as the phase of the wave function minus the semiclassical phase. (This test is carried out in the BBFF $L=0$ case.)\label{LandauZener}}
\end{figure}

Several reasons might explain the discrepancies between the $P$-matrix calculations and the fitted-time-evolution values for the Landau-Zener parameters. One reason for the discrepancies is that the Landau-Zener model as we have implemented it here does not include the coherent phase information of the time-evolution. Thus the phase coherence of the different indistinguishable pathways could affect the final probability distribution, as in the Stueckelberg case. In the analytical solution of the two-level Landau-Zener model for arbitrary initial conditions, the time evolution could be understood as a ``sudden'' jump of phase plus the additional phase accumulated in the simple semiclassical time evolution asymptotically, since the semiclassical approximation accurately describes the slowly varying phase. A numerical test shows that this ``asymptotic'' approximation is only valid in the regime of very slow ramping speed in Fig. \ref{LandauZener}. For each transition, the prediction from the Landau-Zener model has better agreement with the absolute value of the transition probability than with the phase of the transition probability obtained in the numerical calculation. Evidently the LZ model, which predicts the asymptotic behavior of transition probabilities between coupled levels, is more successful in predicting the probability information than the phase information. 
\par
Another possible source of discrepancy is the fact that many levels have avoided crossings that are not sufficiently well isolated, thus invalidating the Landau-Zener model. This error source differs from the first one in the sense that the second considers the effect of several crossings at the same time, while the first error source considers only isolated two-level crossings. 
\par
In addition, considering our calculation's attractive Gaussian model potential, we point out that this potential corresponds to a specific three-body parameter which controls the trimer's size and binding energy. In our calculation, the three body parameter $\kappa$ is defined as $|E_{trimer}|=\frac{\hbar^2\kappa^{2}}{\mu_{trimer}}$, where $E_{trimer}$ is the trimer's binding energy at unitarity without trap, and $\mu_{trimer}$ is the three body reduced mass of the trimer $\mu_{trimer}=(\frac{m_{1}m_{2}m_{3}}{m_{1}+m_{2}+m_{3}})^{1/2}$. For the $^{87}\mbox{Rb}$ and $^{40}\mbox{K}$ Bose-Fermi mixture, $\mu_{trimer}=0.686$, $E_{trimer}=-44.02\hbar\omega$, where the dimensionless parameter has the value $\kappa a_{ho}=5.49$ in our units. For another boson-fermion interaction model potential, if it is within the same $a_{ho}/a_{sc}$ range and has the same dimensionless three body parameter $\kappa a_{ho}$, it will agree with the spectra and dynamics from our calculation near unitarity (unless it is too far from unitarity, in which case the finite range correction of the potential is not negligible.)
\section{Conclusion}
Three and four-body problems remain fundamental and challenging. This study has presented an accurate numerical solution of the spectrum and dynamics of three and four-body Bose-Fermi mixtures in a trap in the vicinity of a Fano-Feshbach resonance. Even though the spectrum presents a rich structure of avoided crossings, a simple Landau-Zener model approximately describes the dynamics of unidirectional ramps and clarifies the relevant characteristic ramping speeds. The spectrum and dynamics are of immediate relevance to optical lattice experiments. These would allow access to physics that cannot be probed in a two-body system, such as trimer-atom interactions and dimer-dimer interactions. Also, the system of two bosons and two spin-polarized fermions in a trap exhibits many of the ingredients of the Fano-Feshbach resonances that form fermionic dipole molecules. In that sense, our results provide a few-body perspective on many-body Bose-Fermi mixture experiments carried out at ultracold temperatures.
\par

\begin{acknowledgments}
We thank Jose D'Incao, Jia Wang, Yujun Wang for discussion about the numerical technique as well as the connection to analytical models. We also thank Shinichi Watanabe and Tomotake Yamakoshi for providing deep insight into of the stochastic phase-space sampling model. This work has been supported in part by the NSF. 
\end{acknowledgments}

\end{document}